# Hybrid Distortion Aggregated Visual Comfort Assessment for Stereoscopic Image Retargeting

Ya Zhou, Zhibo Chen, *Senior Member, IEEE*, and Weiping Li, *Fellow, IEEE*

*Abstract*—Visual comfort is a quite important factor in 3D media service. Few research efforts have been carried out in this area especially in case of 3D content retargeting which may introduce more complicated visual distortions. In this paper, we propose a Hybrid Distortion Aggregated Visual Comfort Assessment (HDA-VCA) scheme for stereoscopic retargeted images (SRI), considering aggregation of hybrid distortions including structure distortion, information loss, binocular incongruity and semantic distortion. Specifically, a Local-SSIM feature is proposed to reflect the local structural distortion of SRI, and information loss is represented by Dual Natural Scene Statistics (D-NSS) feature extracted from the binocular summation and difference channels. Regarding binocular incongruity, visual comfort zone, window violation, binocular rivalry, and accommodation-vergence conflict of human visual system (HVS) are evaluated. Finally, the semantic distortion is represented by the correlation distance of paired feature maps extracted from original stereoscopic image and its retargeted image by using trained deep neural network. We validate the effectiveness of HDA-VCA on published Stereoscopic Image Retargeting Database (SIRD) and two stereoscopic image databases IEEE-SA and NBU 3D-VCA. The results demonstrate HDA-VCA's superior performance in handling hybrid distortions compared to state-of-the-art VCA schemes.

*Index Terms*—Hybrid distortion, stereoscopic image retargeting, visual comfort assessment.

## I. INTRODUCTION

WITH the popularity of various stereoscopic display technologies and applications, such as three-dimensional (3D) television, 3D virtual reality (VR), and augmented reality (AR), stereoscopic content assessment has aroused extensive attention. Different from two-dimensional (2D) images, stereoscopic images contain additional disparity information and delivery a more immersive experience for users. Therefore, stereoscopic image assessment needs to deal with multi-dimensional quality factors including image quality, depth quality, visual comfort, naturalness, sense of presence, etc., which is more complicated than 2D image assessment [1]. Among them, visual comfort is one of the important factors attracting more and more research attentions recently. The term visual comfort refers to the subjective sensation of comfort that can be associated with the viewing of stereoscopic images [1].

Specifically, it directly reflects the physiological experience of human visual system (HVS) and further affects people's desire for immersive multimedia content [2]. Therefore, in the popularization of stereo technology, visual comfort has become an inevitable safety problem that people pay close attention to.

Meanwhile, it is often necessary to process media content with different purposes before delivering them to heterogeneous application scenarios, which is usually called repurposing process in the industry. Therefore, in addition to the visual comfort generated during the production of video content, distortion introduced during repurposing, compression, and transmission may have different effects on the visual comfort experience. In other words, it is not enough to just consider the impact of content, like many existing research works [3]-[5]. Hence, in this paper, we try to study the visual comfort assessment (VCA) in the repurposing process, which is an important part of the multimedia processing chain, and will introduce more influences from depth adjustment, structure modification and information loss.

In the repurposing process, multimedia retargeting techniques have been extensively studied to adapt images/videos to heterogeneous display devices with different screen resolutions, while preserving important content of multimedia. Traditional 2D image retargeting techniques have been illustrated in [6], [7]. Although there are fewer stereoscopic image retargeting (SIR) methods, more and more related researches are devoted to develop a perceptually stereoscopic image retargeting algorithm now, such as stereo warping based on user's quality of experience (QoE) [8], stereo seam carving [9], and stereo multi-operator [10]. Therefore, how to evaluate stereoscopic retargeting technique becomes an important issue. And an excellent SIR method should also consider the factor of visual comfort, which directly affects the user's acceptance of stereo technology. Therefore, this paper focuses on the urgent need for an objective visual comfort metric reflecting human subjective perception assessment, which can not only guide the direction of SIR improvement, but also can help to make the proper choice among several retargeting methods in real applications.

Although many quality assessment studies have been studied over the years, none of them can be applied directly to the VCA for SIR. For example, 2D image quality assessment (IQA) methods do not consider stereo perception [11]-[17], which is one of the significant factors affecting visual comfort. As far as 3D IQA, existing schemes do not consider the characteristics of stereoscopic retargeting operations [18]-[21], where the structural distortion and the information loss introduced by SIR process have an important impact on the end user experience. There are also few 3D image retargeting quality assessment studies [22], [23] which simply consider the image quality and stereo perception quality. However, they lack an analysis of complex binocular perception mechanism in HVS, which is critical for visual comfort evaluation. Similarly, existing 3D VCA research works [3]-[5], [24]-[27] mainly focus on the impacts of raw 3D content, lacking consideration of influences from flexible depth adjustment, structure modification and information loss, all of which are important factors in real 3D content processing chain. Therefore, it is very important and necessary to propose a visual comfort assessment algorithm of 3D image retargeting that comprehensively considers abovementioned hybrid factors.

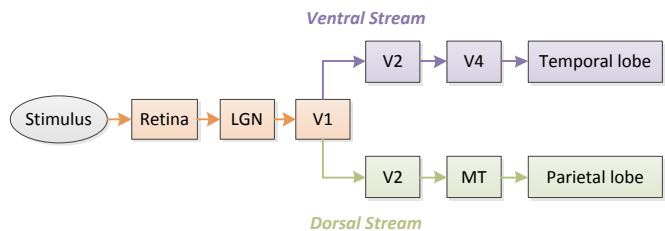

Fig. 1. Illustration of the two-streams hypothesis of human visual system. The ventral stream consists of V1, V2, V4, and temporal lobe areas, and it is implicated with shape recognition and object representation. The dorsal stream consists of V1, V2, MT, and the parietal lobe areas, and it is involved in disparity and motion computations.

This paper proposes a hybrid distortion aggregated visual comfort assessment (HDA-VCA) metric for SIR, which is inspired by the two-streams hypothesis of HVS [28] as shown in Fig. 1. In detail, when the binocular eyes receive stereoscopic visual signals, the photoreceptors on the retina transform the optical signal into an electrical signal, which is then transmitted to the retina ganglion cells where the first receptive field of HVS is formed [4]. The outputs of retina ganglion cells are relayed via the lateral geniculate nucleus (LGN) to primary visual cortex termed as area V1 [29]. The binocular vision signal has been segregated until they are first combined in V1 [30]. Then, there are two separate neural pathways diverging from V1 termed as the ventral and dorsal streams, respectively. The neurons along the ventral stream are mainly implicated with shape recognition and object representation [31]. And the neurons along the dorsal stream are predominantly involved in horizontal disparity primitives, motion computations, such as optical flow [32]. Finally, the semantic perception is formed at the visual center of the cerebral cortex. Therefore, based on the dual pathway sensing of HVS, we design a hybrid distortion aggregated visual comfort assessment metric for stereoscopic image retargeting (VCA-SIR), which was designed and evaluated from the following four aspects.

- Firstly, we propose local structural similarity (Local-SSIM) and dual natural scene statistics (D-NSS) features correspond to the structural arrangements and object information in ventral stream, respectively. Traditional structural similarity (SSIM) measurements cannot be directly used to reflect structural distortion due to the differences in image sizes before and after retargeting. Unlike IR-SSIM [33], which considers visual saliency estimation, the proposed Local-SSIM integrates the advantages of scale-invariant feature transform (SIFT) and SSIM to convert global manipulation to the local computations around key points. Then, the D-NSS is utilized to represent information loss of stereoscopic retargeted images by using the binocular summation and difference channels.
- Secondly, we introduce the measurement of binocular incongruity. In detail, disparity range feature is proposed to reflect the effect of the visual comfort zone of HVS, and perceptual alternation feature is used to illustrate the likelihood of window violation and binocular rivalry in a pair of retargeted images. Meanwhile, we design the disparity intensity distribution feature to represent the adjustment amplitude of accommodation-vergence (A/V) process in HVS. This measurement of binocular incongruity works in concert with the sense of horizontal disparity in dorsal stream and does not consider motion since we only discuss still images here.
- Thirdly, we introduce semantic distortion to represent the higher level stereoscopic perception and understanding process in HVS. Specifically, the semantic distortion is measured by calculating the correlation distance of paired feature maps extracted from original stereoscopic image and its retargeted image by utilizing VGG-16 [34].
- Finally, we verify the effectiveness of the metric on published Stereoscopic Image Retargeting Database (SIRD). The median Pearson line correlation coefficient (PLCC) and Spearman's rank correlation coefficient (SRCC) can reach 0.9326 and 0.9260, respectively. Besides, we also evaluate the performance on IEEE-SA database and NBU 3D-VCA database which both purely consist of stereo images pairs without SIR operation. Experimental results show that HDA-VCA has superior performance in handling multiple distortions compared to the state-of-the-art VCA schemes.

Rest of this paper is organized as follows. Related work is introduced in Section II. In Section III, we introduce the proposed HDA-VCA for SIR metric. The experimental setting, results and analysis are presented in Section IV. And we conclude in Section V.

## II. RELATED WORK

In this section, we mainly introduce relevant quality assessment studies from 2D and 3D perspective. In the 2D field, we first analyze 2D general IQA methods, and then discuss the existing 2D retargeted IQA studies. In the 3D field, in addition to 3D general IQA and 3D retargeted IQA works, we also introduce the research status of 3D VCA which is unique in 3D multimedia processing.

During the last few decades, many successful 2D IQA metrics have been developed, such as the widely used full-reference SIFT flow [11], fast EMD [12], and the no-reference BRISQUE [13], OG-IQA [14]. However, all of them do not take into account the loss of information during SIR process. Also, 2D retargeted image quality assessment (RIQA) has made some progress recently. For example, [15] proposed an objective RIQA metric based on hybrid distortion pooled model named HDPM. A full-reference quality assessment model for image retargeting considering natural scene statistics (NSS) modeling and bi-directional saliency similarity is introduced in [16]. A multiple-level retargeted image quality measure is proposed by [17], which analyzes the aspect ratio similarity, edge group similarity and face block similarity. Although some of them involve information loss and structure distortion measurements, they do not take into account the disparity change introduced by SIR.

Compared with the quality assessment of traditional 2D image, 3D quality assessment involves more perception quality dimensions such as visual comfort and depth quality, in addition to image quality [1]. Chen et al. proposed a full-reference stereoscopic image quality assessment which accounts for binocular rivalry [18]. Lin et al. developed another full-reference 3D IQA metric based on binocular vision model [19]. Liu et al. in [20] introduced a no-reference model driven by binocular spatial activity and reverse saliency. Chen et al. proposed a blind stereoscopic video quality assessment lately, which focused on depth perception and applied its crucial part to the overall 3D video quality assessment [21]. However, the challenges from retargeted images are still not fully studied due to its complexity and multi-distortion characteristics. Few research works have been carried out on 3D RIQA metrics [22], [23]. [22] combines picture quality and stereo perception quality and then uses a polynomial to fit these features. Similar with [22], [23] proposes picture completeness, local distortion and global distortion features in measuring image retargeting quality, and extracts depth similarity, disparity excessiveness features to represent the stereo perception quality. Both of them only evaluate the image quality of stereoscopic retargeted images without consideration of the important factors of visual comfort. And they also lack consideration on binocular perception mechanism of HVS, such as A/V process and binocular rivalry, which are important factors evaluating visual comfort [35].

In recent years, 3D visual comfort quality assessment has attracted much attention. Some VCA models for stereoscopic images have been proposed. For example, in [24], the visual fatigue and visual comfort for 3D HDTV image were studied. [25] measured the disparity distribution and applied it to analyze visual comfort. Also, disparity was utilized to predict the visual comfort of stereoscopic images in [26]. In [27], a visual comfort prediction method for stereoscopic video was introduced, which combined principal component analysis (PCA) and multiple regression. Then, binocular perception mechanism was innovatively considered to predict the 3D visual comfort in [3]. And [4] developed a 3D visual comfort predictor, which extracted coarse features derived from the statistics of binocular disparities and fine features derived by estimating the neural activity associated with the processing of horizontal disparities. Recently, a binocular fusion deep network used to assess the overall degree of visual comfort is devised by [5], which learns binocular characteristics between stereoscopic images. Nevertheless, the above-mentioned methods have been proposed without consideration of impacts from image retargeting, which may have limited application scenarios.

### III. PROPOSED HDA-VCA METRIC

The block diagram of the hybrid distortion aggregated visual comfort assessment (HDA-VCA) for stereoscopic image retargeting (SIR) metric is shown in Fig. 2. It consists of four kinds of measurement. According to [36], some major determining factors for human visual perception on 2D retargeted images, i.e., global structural distortion, local region distortion, and loss of salient information, are mentioned. Firstly, the proposed Local-SSIM feature is utilized to represent the local region distortion, i.e., the local structural distortion. Secondly, the loss of salient information is covered by D-NSS feature. Thirdly, in the binocular incongruity measurement, inspired by the peculiarity and the binocular vision mechanism of HVS, disparity range feature, perceptual alternation feature and disparity intensity distribution feature are designed specific to different perceptual factors related to visual comfort in HVS. Fourthly, in order to describe semantic distortion, which indirectly reflects the global structural distortion, the semantic distortion feature is introduced by utilizing the popular deep convolution network VGG-16. Finally, all the features are pooled into Support Vector Regression (SVR) model to generate the final predicted visual comfort scores.

*A. Local Structural Distortion Measurement*

Inspired by the earlier 2D image retargeting quality assessment work [16], [37], [38], we also consider structural distortion in our metric. Most SIR operations still introduce this type of distortion while also producing some distinct distortions, such as stereo seam carving, as shown in Fig. 3. Comparing (a) and (b) in Fig. 3, we can observe that there are many twisty areas. Since seam carving is an operation which tends to delete the seam whose certain defined energy is the minimum value, the deleted seam may be not a straight pixel line. Thus, the primary structure between the adjacent pixels will be destroyed. The local structural distortion is considered in the section, while the global structural distortion will be analyzed as a kind of semantic distortion in this paper.

The most accepted structural distortion measurement SSIM cannot be directly applied to stereoscopic retargeted images due to the different image size between the original image and the retargeted image [39]. There have been some works like SIFT flow [11] and IR-SSIM [33] to deal with this problem. But SIFT flow is used to find the best matched result for one image while it does not measure the structural distortion. And IR-SSIM utilizes an automatic saliency prediction approach to generate a saliency map and then conduct a spatially varying weighting for SSIM map, whose performance relies on the saliency detector. However, saliency areas may change before and after retargeting owing to the retargeting operations. For example, as shown in Fig. 3 (a), the saliency area is the central square building, while as for Fig.3 (b), the deformed areas in the background attract more attention since the human eyes are very sensitive to structural deformation. Therefore, it is important to pay more attention to preserve the structure of the areas around key structure points in images, not just the so-called salient areas. Therefore, we utilize SIFT to find and match the key points [40], and then use SSIM to measure the local structural distortion. We name the derived feature as Local Structural SIMilarity (Local-SSIM).

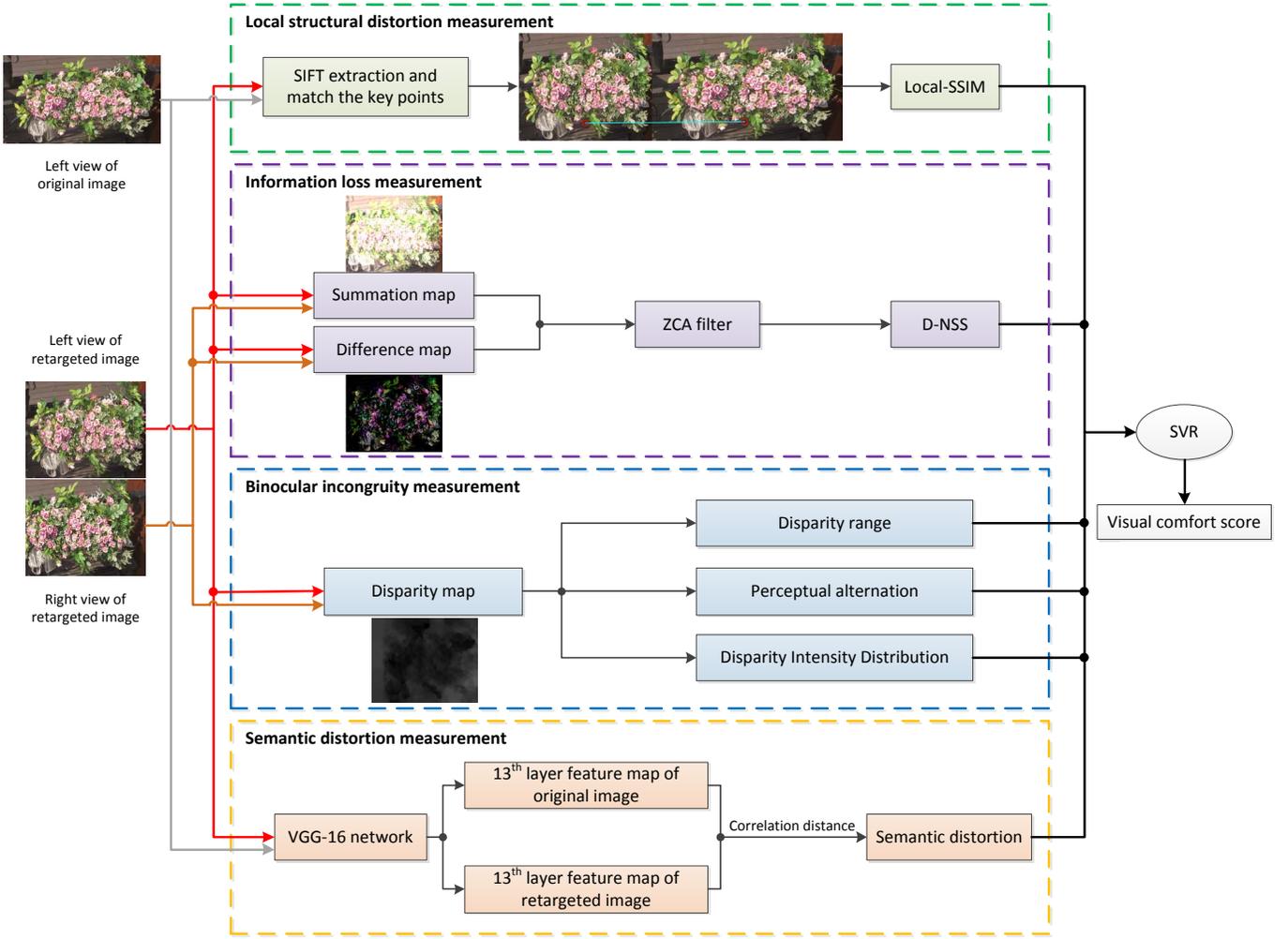

Fig. 2. Block diagram of the Hybrid Distortion Aggregated Visual Comfort Assessment (HDA-VCA) for Stereoscopic Image Retargeting (SIR) scheme. The top green dotted line block is the flow chart of local structural distortion measurement. The next purple dotted line block represents the process of information loss measurement. The middle blue dotted line block is the flow chart of binocular incongruity measurement containing three key aspects: disparity range, perceptual alternation, and disparity intensity distribution features. And the bottom orange dotted line block represents the process of semantic distortion measurement.

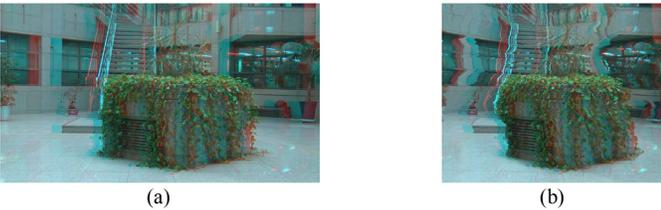

Fig. 3. Illustration of structural distortion and information loss in stereo seam carving. (a) A pair of stereoscopic original image; (b) A pair of correspondingly stereoscopic seam carving image. Both of them are displayed by overlapping the left view image and the right view image.

Firstly, we obtain the gray-scale maps of the original left view image and the retargeted left view image as the processing objects, which are denoted as $O_l$ and $R_l$, respectively. The key points $\{R_l(m,n)\}$ are extracted from the retargeted left view image $R_l$ by using SIFT. In the same way, the key points $\{O_l(p,q)\}$ are also extracted from the original left view image $O_l$. We further screen out the matching key points $\{O_l(m',n')\}$ among $\{O_l(p,q)\}$ based on $\{R_l(m,n)\}$. Secondly, square areas with the center of each point in $\{R_l(m,n)\}$ and $\{O_l(m',n')\}$ are defined. Then, based on two corresponding square areas as shown in Fig. 4, the local SSIM value can be calculated as follows [39]:

$$SSIM(x,y) = \frac{(2\mu_x\mu_y + C_1)(2\sigma_{xy} + C_2)}{(\mu_x^2 + \mu_y^2 + C_1)(\sigma_x^2 + \sigma_y^2 + C_2)} \quad (1)$$

where

$$\mu_x = \sum_{i=1}^{N} w_i x_i \text{ and } \mu_y = \sum_{i=1}^{N} w_i y_i \quad (2)$$

$$\sigma_x = \sqrt{\sum_{i=1}^{N} w_i (x_i - \mu_x)^2} \text{ and } \sigma_y = \sqrt{\sum_{i=1}^{N} w_i (y_i - \mu_y)^2} \quad (3)$$

$$\sigma_{xy} = \sum_{i=1}^{N} w_i (x_i - \mu_x)(y_i - \mu_y) \quad (4)$$

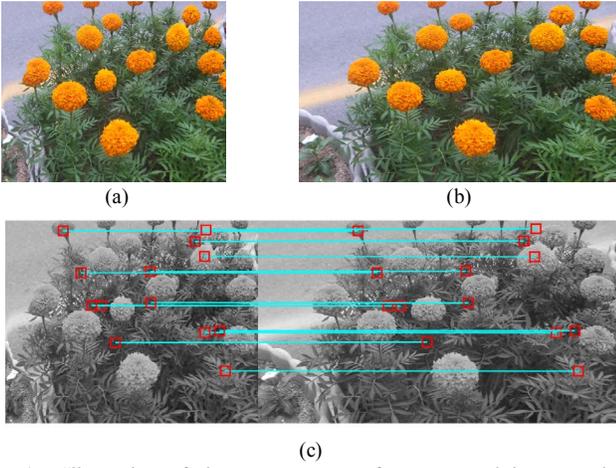

Fig. 4. Illustration of the square areas of a retargeted image and the corresponding square areas of its original image. (a) The left view image of stereoscopic retargeted image; (b) The left view image of stereoscopic original image; (c) The matched sample of square areas between the stereoscopic retargeted image and its original image.

---

**Algorithm 1** Local Structural Distortion Measurement

**Input:** Luminance maps of retargeted left view image and its original left view $R_l$, $O_l$

**Output:** Local-SSIM feature

1: $\{R_l(m, n)\} \leftarrow$ SIFT key points extraction from $R_l$
2: $\{O_l(p, q)\} \leftarrow$ SIFT key points extraction from $O_l$
3: Screen out the matched point $\{O_l(m', n')\}$ from $\{O_l(p, q)\}$ based on $\{R_l(m, n)\}$
4: **for** each pair of $(m, n)$ and $(m', n')$ **do**
5: $\quad A_R = R_l(m-16{:}m+16, n-16{:}n+16)$
6: $\quad A_O = O_l(m'-16{:}m'+16, n'-16{:}n'+16)$
7: $\quad$ Generate $\widehat{SSIM} \leftarrow A_R, A_O$ by Eq. (1,2,3,4)
8: **end for**
9: $f_{Local-SSIM} = mean(\widehat{SSIM})$
10: **return** $f_{Local-SSIM}$

---

Here, $w$ is an 11*11 Gaussian window and $(x, y)$ is spatial indices. Thirdly, the average of all the SSIM values is served as Local-SSIM feature of the stereo pair. Detail of the algorithm can be referred to the pseudocode in Algorithm 1.

*B. Information Loss Measurement*

NSS method has been applied to many quality assessment tasks [41], [42]. In IQA domain, NSS measurement utilizes the principle that natural images possess certain regular statistic properties that are measurably modified by the presence of certain distortions [16].

Some stereoscopic image retargeting methods may only operate on certain part of images, for example, as shown in Fig. 5 (a) and (b), stereo cropping just crops the left and right parts of images which is not a uniform operation. Since the number of pixels in the gray range of cropped areas declines more than that of the non-cropped areas, as the red curve shown in Fig. 5 (d) and (e), the gray-level distribution of these retargeted images has less similarity with the distribution before retargeting, even after normalization.

There are also some stereoscopic retargeting methods which operate images globally, such as stereo scaling which is shown in Fig. 5 (c). The gray-level distribution of these images is closer to that before retargeting, as the orange curve shown in Fig. 5 (d) and (e). We can regard these retargeting methods as relatively uniform operations. Assuming that the size of retargeted images is fixed, the information integrity under inhomogeneous loss is worse than that under loss sharing. In this way, information integrality can be reflected by the gray-level distribution of retargeted images.

In this paper, the dual channel maps, i.e., the summation map and the difference map, are obtained to model the binocular fusion and the disparity perception of HVS in a preliminary way, as below:

$$R^+ = R_l + R_r \quad (5)$$
$$R^- = R_l - R_r \quad (6)$$

where $R_l$ and $R_r$ are the retargeted left and right view images, respectively. The difference map can be regarded as the expression of salient information. Because the depth perception is stronger when the difference is greater, and the more attention the area can catch. Then, the zero-phase component analysis (ZCA) whitening filter is applied to reduce the correlation among pixels as:

$$Z^+ = ZCA(R^+) \text{ and } Z^- = ZCA(R^-) \quad (7)$$

The gray-level distribution of summation map and difference map after ZCA are shown in Fig. 5 (f) and (g), respectively. Inspired by [13], the mean subtracted contrast normalized (MSCN) coefficients are computed on the two maps. And neighboring MSCN coefficients along four orientations between neighboring pixels and two scales are also considered as the setting in [13]. Then we use the asymmetric generalized Gaussian distribution (AGGD) with zero mean value to fit the distribution as given by [43]:

$$f(x; \lambda, \sigma_l^2, \sigma_r^2) = \begin{cases} \dfrac{\lambda}{(\rho_l + \rho_r)\Gamma(1/\lambda)} e^{-\left(\frac{-x}{\rho_l}\right)^\lambda} & x < 0 \\ \dfrac{\lambda}{(\rho_l + \rho_r)\Gamma(1/\lambda)} e^{-\left(\frac{x}{\rho_r}\right)^\lambda} & x \geq 0 \end{cases} \quad (8)$$

where $\lambda$, $\sigma_l^2$, $\sigma_r^2$ are the shape parameter and the scale parameters of the left and right sides respectively, and

$$\rho_l = \sigma_l \sqrt{\dfrac{\Gamma(1/\lambda)}{\Gamma(3/\lambda)}} \text{ and } \rho_r = \sigma_r \sqrt{\dfrac{\Gamma(1/\lambda)}{\Gamma(3/\lambda)}} \quad (9)$$

where $\Gamma(\cdot)$ is the gamma function:

$$\Gamma(t) = \int_0^\infty r^{t-1} e^{-r} dr \quad t > 0 \quad (10)$$

Besides, the parameter $\eta$ is computed to reflect the difference of two sides as [21]:

$$\eta = (\rho_l - \rho_r) \dfrac{\Gamma(2/\lambda)}{\Gamma(1/\lambda)} \quad (11)$$

Thus, the 64-dimensional parameters $(\eta, \lambda, \sigma_l^2, \sigma_r^2)$ of the best AGGD fit for the gray-level distribution of summation map and difference map are served as the feature representing the information loss, which is named Dual Natural Scene Statistics (D-NSS).

The pseudocode of the information loss measurement is

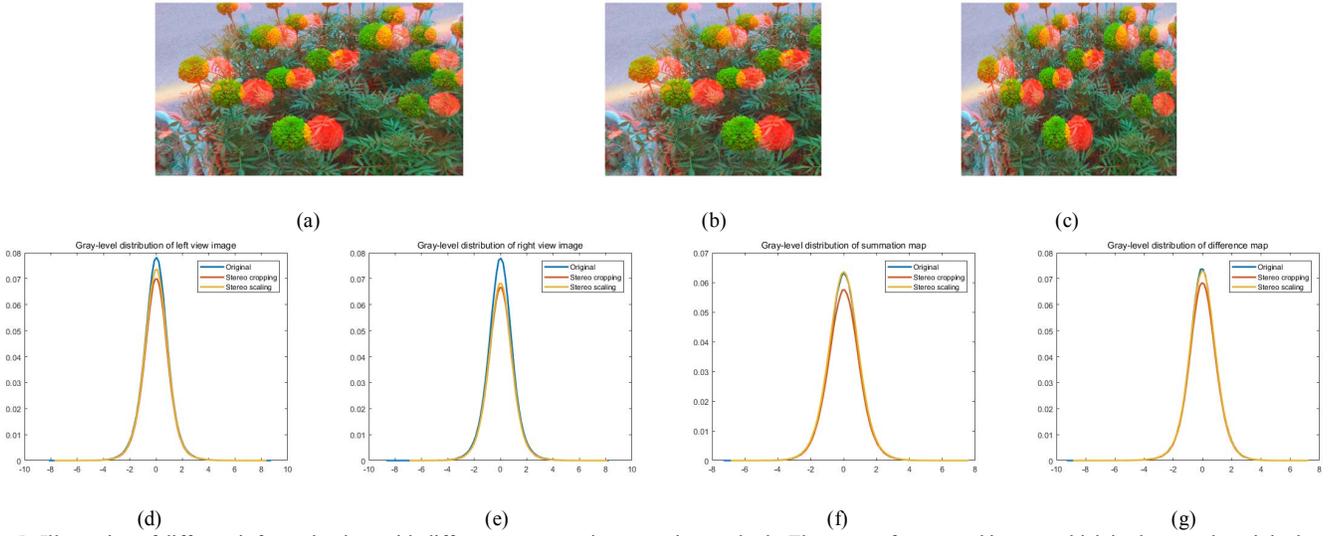

Fig. 5. Illustration of different information loss with different stereoscopic retargeting methods. The curve of retargeted images which is closer to the original curve represents the integrity of information is higher. (a) Stereoscopic original image; (b) Correspondingly stereo cropping image; (c) Correspondingly stereo scaling image; (d) Gray-level distribution of left view images after ZCA; (e) Gray-level distribution of right view images after ZCA; (f) Gray-level distribution of summation map after ZCA; (g) Gray-level distribution of difference map after ZCA.

---

**Algorithm 2** Information Loss Measurement
**Input:** Luminance maps of left and right view images, i.e., $R_l, R_r$
**Output:** D-NSS feature
1: $R^+ = R_l + R_r \leftarrow$ *summation map*
2: $R^- = R_l - R_r \leftarrow$ *difference map*
3: Generate ZCA whitening filtered maps
   $Z^+ = ZCA(R^+), Z^- = ZCA(R^-)$
4: Generate MSCN coefficient maps
   $MSCN^+ = MSCN(Z^+), MSCN^- = MSCN(Z^-)$
5: **for** each orientation $o$ and each scale $s$ **do**
6:    Generate neighboring MSCN coefficient map
      $MSCN^+_{o,s}, MSCN^-_{o,s}$
7:    Quantify the statistical distribution of $MSCN^+_{o,s}$ by
      Eq. (8,9,10,11):
8:    $(\eta_{1,o,s}, \lambda_{1,o,s}, \sigma^2_{l1,o,s}, \sigma^2_{r1,o,s}) \leftarrow AGGD\ model$
9:    Quantify the statistical distribution of $MSCN^-_{o,s}$ by
      Eq. (8,9,10,11):
10:   $(\eta_{2,o,s}, \lambda_{2,o,s}, \sigma^2_{l2,o,s}, \sigma^2_{r2,o,s}) \leftarrow AGGD\ model$
11:   $f_{o,s} = (\eta_{1,o,s}, \lambda_{1,o,s}, \sigma^2_{l1,o,s}, \sigma^2_{r1,o,s}, \eta_{2,o,s}, \lambda_{2,o,s}, \sigma^2_{l2,o,s}, \sigma^2_{r2,o,s})$
12:   Update $f_{D-NSS} \leftarrow [f_{D-NSS}, f_{o,s}]$
13: **end for**
14: **return** $f_{D-NSS}$

presented in Algorithm 2.

### C. Binocular Incongruity Measurement

According to the characteristics of SIR, the SIR operations can adjust the disparity of stereo pairs to different degrees, such as stereo seam carving, stereo scaling, and stereo multi-operator. Therefore, the original disparity of comfort viewing stereo pair can be changed to be outside of visual comfort zone of HVS, and cause various uncomfortable binocular perception. In binocular incongruity measurement, we consider several impacts on visual comfort, namely visual comfort zone, window violation, binocular rivalry, and accommodation-vergence (A/V) conflict. Accordingly the features of disparity range, perceptual alternation, and disparity intensity distribution are designed in this paper.

*1) Disparity Range feature*

When the left and right view images are presented to each eye respectively, there are several binocular perceptual modes, i.e., binocular fusion, binocular rivalry, and binocular suppression. As the disparity increases, the above-mentioned three visual modes will occur in an orderly manner. Binocular fusion refers to a kind of visual comfort situation that human eyes can fuse left and right view images normally and then form a stereo or depth perception. Conversely, binocular rivalry and binocular suppression are two kinds of uncomfortable vision status, which will be introduced in the perceptual alternation feature.

When the disparity range of a stereo pair is large enough, it is difficult for the eyes to fuse the left view and the right view images, and the binocular fusion state is broken. Meanwhile, the A/V conflict occurs [35], because the adjustment process of accommodation tends to adjust the focus of the eyes on the screen, and the vergence process tends to adjust the eyes to focus on the stereoscopic objects and perceive the virtual depth. The two processes cannot reach a balance. Then, the viewers will feel tired and uncomfortable. How to find the critical value of disparity range and measure the possibility of broking binocular fusion becomes a crucial issue. Based on the earlier visual physiology research [44], visual comfort zone is defined to represent the acceptable disparity range, i.e., ±1° in angular disparity. Inspired by this, we use the weighted difference between disparity range and visual comfort zone as a measure of the aforementioned possibility, which is defined as:

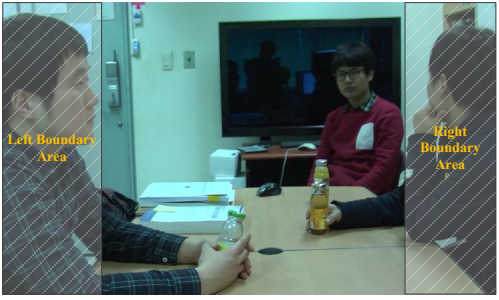

Fig. 6. Illustration of the definition of boundary area. The left and right gray planes denote the left and right boundary areas separately.

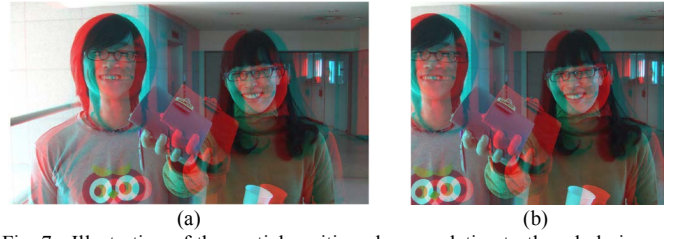

(a)　　　　　　　　　　　(b)

Fig. 7. Illustration of the spatial position change relative to the whole image introduced by stereoscopic cropping. (a) Stereoscopic original image; (b) Stereo cropping image. The man is originally located near the center of the image, but after stereo cropping, he is located at the boundary of the image.

$$R = \alpha \frac{d_{\min} - d_l}{d_l} + \beta \frac{d_{\max} - d_u}{d_u} \quad (12)$$

$$\alpha + \beta = 1 \quad (13)$$

where $[d_l, d_u]$ is the disparity range of each pair of stereoscopic retargeted images, and $[d_{min}, d_{max}]$ is the range of visual comfort zone, which is [-79.55, 79.55] pixel computed on our experimental setting. $\alpha$ and $\beta$ are the penalty factors of the minimum and maximum disparity values beyond the comfort zone respectively. According to [4], the MT neurons of HVS prefer the crossed disparity. Namely, human eyes are more sensitive to the discomfort caused by crossed disparity than uncrossed disparity. Hence, we set $\alpha = 0.6$ and $\beta = 0.4$ in the metric.

*2) Perceptual Alternation feature*

Some stereoscopic retargeting operations may change the spatial position of foreground objects to the boundary area, as illustrated in Fig. 6 and Fig. 7. In this case, the disparity in the boundary area is larger than that in other areas since foreground objects usually have relatively larger disparity. Then, the larger disparity in the boundary area is more likely to induce window violation, due to the limitation of visual field [45]. Intuitively, the viewers perceive a glint formed by the image content and a black stripe near the boundary, which looks like the shake of display window and is called as window violation. Furthermore, analyzing from the perspective of information transfer, the foreground objects located in the boundary area sometimes cannot be fully displayed, which tends to produce an incomplete sensation of content.

At the same time, for the entire image, we consider whether binocular rivalry occurs. In detail, when left and right view stimulus are completely different, human perception alters between left and right view visual signals, which is known as binocular rivalry. While binocular suppression can be regarded as the extreme case of binocular rivalry, where two eyes are only able to perceive one view image with larger energy and have no stereo sensation, like 2D stimulus. Some stereoscopic retargeting operators, like stereo cropping and stereo scaling, are depth-adjusted, and they can bring in large disparity which makes two view images look very different, even though they have the similar content. Under this circumstances, binocular rivalry occurs during the viewing. Hence, it is necessary to analyze the discomfort effect of binocular rivalry in this feature.

First, we calculate the average disparity of the boundary area to represent the probability of window violation and the completeness of image content. Let the size of the image is $m * n$, and the gray and disparity values on the $i^{th}$ row and $j^{th}$ column of an image are $v_{ij}$ and $d_{ij}$, respectively. The left and right boundaries of a pair of stereo image are defined as follow:

$$b_l = \frac{\sum_{i=1}^{n} d_{i1}}{n} \quad \text{and} \quad b_r = \frac{\sum_{i=1}^{n} d_{im}}{n} \quad (14)$$

Then, the average disparity of the left and right boundary areas are calculated as:

$$A_l = \frac{\sum_{j=1}^{b_l}\sum_{i=1}^{n} d_{ij}}{b_l \cdot n} \quad \text{and} \quad A_r = \frac{\sum_{j=m-b_r}^{m}\sum_{i=1}^{n} d_{ij}}{b_r \cdot n} \quad (15)$$

Second, motivated by existing visual study [46], we utilize the variance of gray values to define the energy of each image as:

$$E = \frac{\sum_{j=1}^{m}\sum_{i=1}^{n}(v_{ij} - \bar{v})^2}{m \cdot n} \quad (16)$$

where $\bar{v}$ is the mean of gray values in the image. If the difference between the energy of left and right view images is large enough, binocular rivalry is considered to occur. Therefore, we calculate the ratio of the two energy values to reflect the possibility of binocular rivalry which is defined as:

$$R_E = \frac{E_l}{E_r} \quad (17)$$

where $E_l$ and $E_r$ are the energy of left and right view images respectively. Then, the three variates $A_l$, $A_r$ and $R_E$, constitute the perceptual alternation feature.

*3) Disparity Intensity Distribution feature*

Accommodation-vergence (A/V) conflict is known as the main factor causing visual discomfort [47]. As we can see from the disparity range feature, it reflects the possibility of A/V conflict. However, it is not enough to analyze the effect of A/V process only from the possibility, since some stereoscopic image retargeting operators may squeeze the disparity distribution, such as stereo scaling. Imagining that if the disparity fluctuation from one local area to another local area is sufficiently obvious, the steady state between accommodation and vergence processes has to be broken and re-established. Also, the new balance has a significant difference from the previous one due to the changed disparity [48]. At this time, viewers usually need a short time to adapt to the newly changed virtual depth by re-focusing, which is done by A/V adjustment.

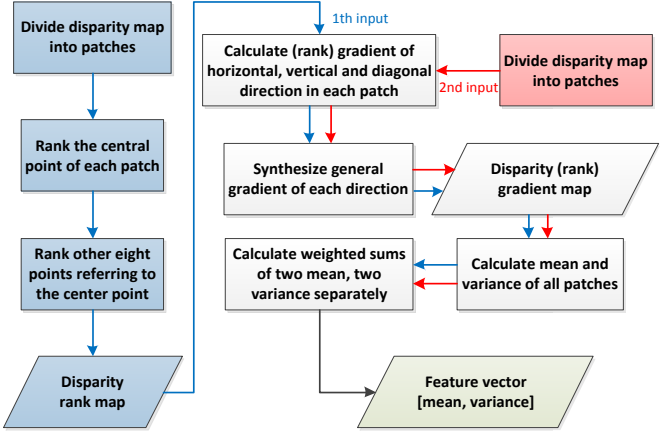

Fig. 8. Diagram of disparity intensity distribution feature.

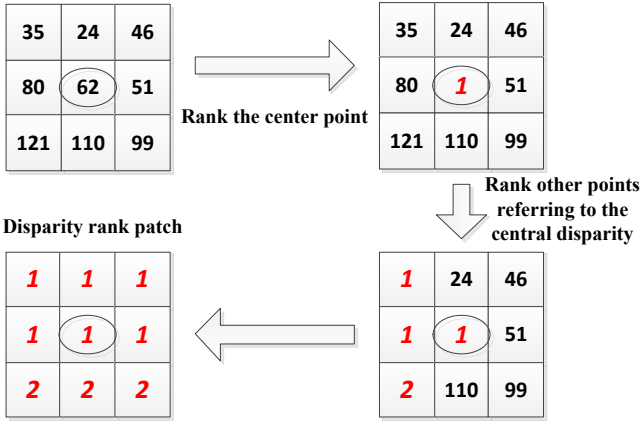

Fig. 9. Illustration of ranking the pixels in a patch based on Just Noticeable Depth Difference (JNDD).

Intuitively, when viewers just see the stereoscopic images with many acute disparity fluctuations at the first sight, they usually feel dazzled. Hence, we propose the disparity intensity distribution feature to reflect the adjustment intensity of A/V process.

The diagram of disparity intensity distribution feature is illustrated in Fig. 8. As the blue and red arrow lines shown in the figure, there are two branches of feature extraction, which are based and not based on ranking respectively. The introduction of ranking is inspired by the grading idea of Just Noticeable Depth Difference (JNDD). JNDD denotes that the perceived disparity difference is limited when seeing stereoscopic images [49]. In the first branch, we first divide the entire disparity map $D$ into patches, whose size is 3*3 in our metric, and then classify the central pixel of each patch into one of the four bins, according to the interval of JNDD which is defined as [49]:

$$D_{JND} = \begin{cases} 21, & \text{if } 0 < |d_{ij}| < 64 : Bin1 \\ 19, & \text{if } 64 \leq |d_{ij}| < 128 : Bin2 \\ 18, & \text{if } 128 \leq |d_{ij}| < 192 : Bin3 \\ 20, & \text{if } 192 \leq |d_{ij}| < 255 : Bin4 \end{cases} \quad (18)$$

where $D_{JND}$ represents the minimum disparity difference value which can be perceived by human eyes. In the same way, as demonstrated in Fig. 9, the other eight pixels in the patch are

**Algorithm 3** Binocular Incongruity Measurement

**Input:** Left and right view images $R_l$ and $R_r$
**Output:** Disparity range feature $f_{DR}$, perceptual alternation feature $f_{PA}$, and disparity intensity distribution feature $f_{DID}$
1: $D \leftarrow$ disparity map from $R_l$ and $R_r$
2: $d_l = min(D)$ and $d_u = max(D)$
3: $d_{min}, d_{max} \leftarrow$ disparity range of visual comfort zone
4: $f_{DR} = \alpha \frac{d_{min} - d_l}{d_l} + \beta \frac{d_{max} - d_u}{d_u}$
5: Generate $A_l, A_r \leftarrow D$ by Eq. (14,15)
6: Generate $R_E \leftarrow R_l, R_r$ by Eq. (16,17)
7: $f_{PA} = [A_l, A_r, R_E]$
8: Divide $D$ into patches $D(i-1:i+1, j-1:j+1)$
9: **for** $i = 1 \rightarrow I, j = 1 \rightarrow J$ **do**
10:     Quantify $D(i-1:i+1, j-1:j+1)$ by Eq. (18):
11:     $D_{rank}(i-1:i+1, j-1:j+1) \leftarrow D(i-1:i+1, i-1:j+1)$
12:     Generate $G_h^{rank}, G_v^{rank}, G_d^{rank} \leftarrow D_{rank}$ by Eq. (19, 20,21)
13:     $G^{rank}(i,j) = \sqrt{G_h^{rank^2} + G_v^{rank^2} + G_d^{rank^2}}$
14:     Generate $G_h, G_v, G_d \leftarrow D$ by Eq. (19,20,21)
15:     $G^{non}(i,j) = \sqrt{G_h^2 + G_v^2 + G_d^2}$
16: **end for**
17: $m^{rank} = mean(G^{rank})$ and $v^{rank} = var(G^{rank})$
18: $m^{non} = mean(G^{non})$ and $v^{non} = var(G^{non})$
19: $m = wm^{rank} + (1-w)m^{non}$
20: $v = wv^{rank} + (1-w)v^{non}$
21: $f_{DID} = [m, v]$
22: **return** $f_{DR}, f_{PA}$ and $f_{DID}$

also classified into the four bins based on the difference between their values and the central disparity. The same rank numbers in adjacent pixels represent that they are perceived as the same disparity level by human eyes. Thus, we obtain the disparity rank map $D_{rank}$, and further calculate the gradients of horizontal, vertical and diagonal directions in its each patch as:

$$G_h = \sqrt{\frac{\sum_{i=-1}^{1}\sum_{j=0}^{1}[D_{rank}(2+i, 2+j) - D_{rank}(2+i, 2+j-1)]^2}{3*2}} \quad (19)$$

$$G_v = \sqrt{\frac{\sum_{i=0}^{1}\sum_{j=-1}^{1}[D_{rank}(2+i, 2+j) - D_{rank}(2+i-1, 2+j)]^2}{3*2}} \quad (20)$$

$$G_d = \sqrt{\frac{\sum_{i=0}^{1}\sum_{j=0}^{1}[D_{rank}(2+i, 2+j) - D_{rank}(2+i-1, 2+j-1)]^2}{2*2}}$$
$$+ \sqrt{\frac{\sum_{i=-1}^{0}\sum_{j=0}^{1}[D_{rank}(2+i, 2+j) - D_{rank}(2+i+1, 2+j-1)]^2}{2*2}} \quad (21)$$

where $G_h$, $G_v$ and $G_d$ are horizontal, vertical and diagonal gradients separately. The general gradient is then synthesized.

Finally, the mean and variance of general gradients of all patches are computed. Meanwhile, in the second branch, the computation process keeps the same only without ranking the pixel. The weighted sums of mean and variance in two branches are obtained separately, which serve as the disparity intensity distribution feature.

The pseudocode of the binocular incongruity measurement is presented in Algorithm 3.

*D. Semantic Distortion Measurement*

Semantic distortion in image retargeting is rarely considered by previous works [15]-[17], [22], [23], though it often happens when objects or content are deformed by image retargeting methods. For example, as shown in Fig. 10 (a) and (b), stereo seam carving deforms the contours of human and background in the image. Since deep convolutional neural networks (CNN) came to the top on the image classification task, CNN has shown it superior ability at extracting high-level features on various fields [50], and its progressive receptive field extension is highly similar with HVS [51]. Based on these, the semantic distortion feature derived from CNN can reflect the perceptive distortion of HVS caused by SIR more directly.

One of the most common networks used to extract the features of images is VGG-16 [34]. Hence, in our metric, we first use left view images as the inputs and then obtain the feature maps of $13^{th}$ convolutional layer in VGG-16, i.e., the last convolutional layer. Feature map is a kind of high-level semantic expression, representing the abstract of certain corresponding area in the input image [52], as the first three channels of feature maps shown in Fig. 10 (c) and (d). In order to measure the similarity between the feature maps of original image and retargeted image, we hereby employ correlation distance to reflect it. The feature maps of original left view images and retargeted left view images are denoted by $F_o$ and $F_r$, respectively. We use the mean of each channel to represent this channel's feature map and denote the obtained results as $\overline{F}_o$ and $\overline{F}_r$. Therefore, the two variates have the same size which is equal to the number of channels $t$. Then, the correlation distance is given as below:

$$C = 1 - \frac{(\overline{F}_o - \widehat{F}_o)(\overline{F}_r - \widehat{F}_r)}{\sqrt{\sum_{i=1}^{t}\left[(\overline{F}_o(i) - \widehat{F}_o)\right]^2}\sqrt{\sum_{i=1}^{t}\left[(\overline{F}_r(i) - \widehat{F}_r)\right]^2}} \quad (22)$$

where $\widehat{F}_o$ and $\widehat{F}_r$ are the mean of $\overline{F}_o$ and $\overline{F}_r$. Variate $C$ serves as the semantic distortion feature.

In this way, we combine the neural network based feature with the traditional hand-craft features and form a complete hybrid distortion aggregated visual comfort assessment (HDA-VCA) metric.

## IV. PERFORMANCE EVALUATION

Due to the lack of public subjective stereoscopic image retargeting databases, as shown in Fig. 11, we have established a Stereoscopic Image Retargeting Database (SIRD) in our previous work [53], which is available online for public research usage [54]. In the SIRD, there are 100 source images and 400 stereoscopic retargeted images, and four typically stereoscopic retargeting methods are involved, i.e., stereo cropping, stereo seam carving, stereo scaling, and stereo multi-operator. The resolution of source images is 1920*1080 pixels, and the resolution of retargeted images is 1344*1080, that is, the shrinking ratio in the column direction is 0.7.

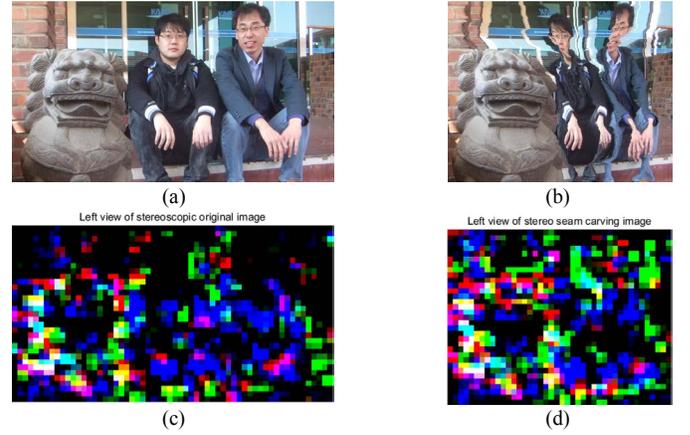

Fig. 10. An example shows the feature map is a high-level abstract of its input image. (a) The left view of stereoscopic original image; (b) The left view of stereo seam carving image; (c) The first three channels' visualization of feature map of (a); (d) The first three channels' visualization of feature map of (b).

Besides, in order to illustrate the universality and versatility of HDA-VCA metric, we also test the performance on the public IEEE-SA database and NBU 3D-VCA database, respectively. IEEE-SA database purely consists of 800 general stereoscopic images without SIR and provides visual comfort scores [55]. And the difference of each pair of stereo image only lies in the image content and its disparity. The resolution is 1920*1080. In the recent years, many visual comfort assessment algorithms for stereoscopic image have been proposed based on the IEEE-SA database, and we will compare them with the proposed metric in Section IV. B. However, NBU 3D-VCA database also purely consists of 200 stereoscopic images with 1920*1080 resolution and the corresponding visual comfort scores [56]. The performance and comparison on the NBU 3D-VCA is presented in Section IV. C.

For the above three databases, we randomly divide each database into 80% for training and 20% for testing and use the associated mean opinion score (MOS) values of visual comfort as the labels. 100 iterations of cross validation are performed on each database. Then, the mean Pearson line correlation coefficient (PLCC), Spearman rank correlation coefficient (SRCC), Kendall rank-order correlation coefficient (KRCC) and root mean square error (RMSE) values between the predicted scores and the subjective MOS values are utilized to represent the final performance.

*A. Performance and Comparison on SIRD Database*

3D visual comfort analysis on heterogeneous devices is an important work for future 3D and immersive visual experiencing assessment. Although there have been some 3D VCA solutions, as the characteristics of stereoscopic retargeting operations, none of them can be directly applied for 3D image retargeting scenarios. Therefore, we validate the VCA denotes visual comfort assessment. IRQA denotes image retargeting quality assessment. And IQA denotes image quality assessment effectiveness from various quality assessment perspective by comparing our proposed HDA-VCA metric with some state-of-the-art methods on SIRD database.

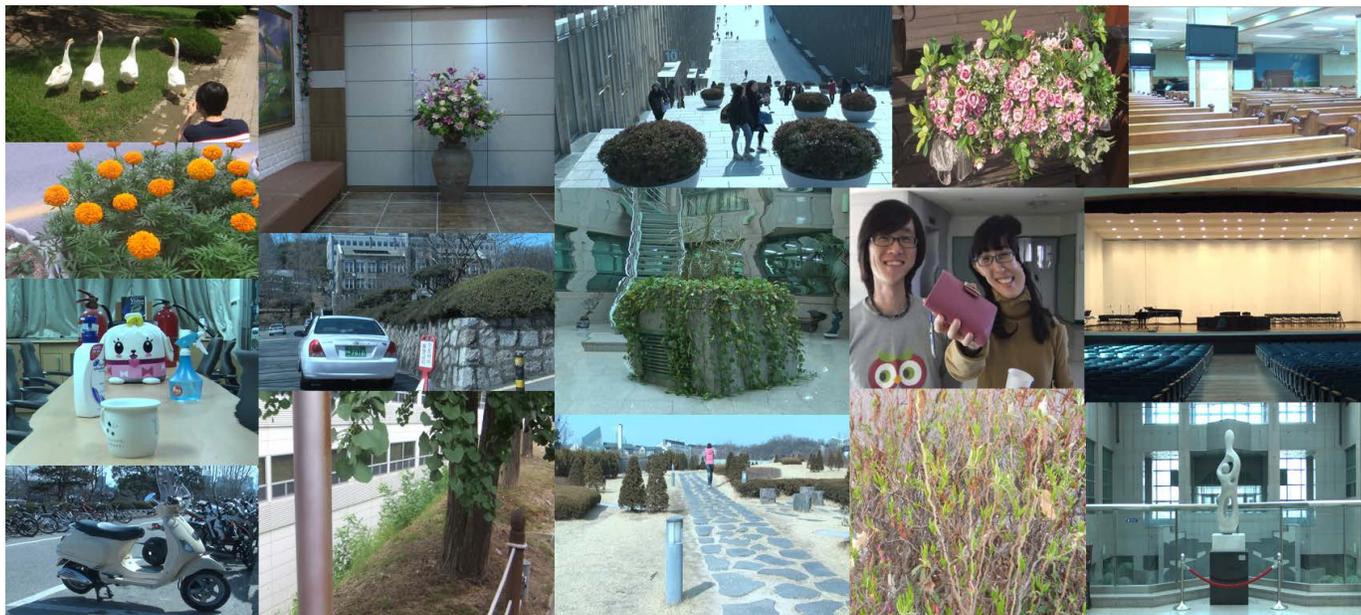

Fig. 11. Example images from Stereoscopic Image Retargeting Database (SIRD). There are four typically stereoscopic image retargeting operations involved in SIRD database, i.e., stereo cropping, stereo seam carving, stereo scaling, and stereo multi-operator.

TABLE I
PERFORMANCE AND COMPARISON ON SIRD DATABASE

| | Metric | PLCC | SRCC | KRCC | RMSE |
|---|---|---|---|---|---|
| 3D VCA | 3DAVM [3] | 0.5737 | 0.5699 | 0.3996 | 0.8705 |
| | Park et al. [4] | **0.5768** | **0.5847** | **0.4109** | **0.8781** |
| 3D IRQA | SRIQA [23] | **0.3991** | **0.3960** | **0.2747** | **0.8957** |
| | MJ3DFR [18] | 0.4310 | 0.3893 | 0.2493 | 0.8615 |
| 3D IQA | BSVQE [21] | **0.8715** | **0.8451** | **0.6552** | **0.5162** |
| | SINQ [20] | 0.8560 | 0.8590 | 0.6780 | 0.5522 |
| | Lin et al. [19] | 0.6696 | 0.6668 | 0.4968 | 0.7955 |
| 2D IRQA | BNSSD [16] | **0.8334** | **0.8298** | **0.6377** | **0.5863** |
| | HDPM [15] | 0.7757 | 0.7607 | 0.5745 | 0.6767 |
| | MLF [17] | 0.1629 | 0.1349 | 0.1150 | 3.3605 |
| 2D IQA | Fast EMD [12] | 0.4892 | 0.4651 | 0.3284 | 0.9384 |
| | SIFT flow [11] | 0.2054 | 0.1625 | 0.1105 | 1.0559 |
| | BRISQUE [13] | **0.8789** | **0.8573** | **0.6711** | **0.5046** |
| | OG-IQA [14] | 0.7786 | 0.7722 | 0.5777 | 0.6601 |
| | Proposed | **0.9326** | **0.9260** | **0.7722** | **0.3772** |

As can be seen from Table I, because there are few public VCA algorithms, without loss of generality, we firstly employ two representative VCA for ordinary stereoscopic image models to contrast on SIRD database, which both can obtain the superior performance on IEEE-SA database as presented in Table II [3], [4]. Secondly, considering the relevance to the VCA-SIR task, we reproduce one of the only two image quality assessment for SIR metrics whose proposer are the same person [23]. It is not hard to find from the low correlation values that although IQA has certain connection with VCA, visual comfort cannot be fully represented by image quality. In other words, there is a distinction between VCA and IQA, and VCA problems should also be taken seriously like IQA researches. Thirdly, the proposed metric is compared with several stereoscopic quality assessment algorithms, i.e., MJ3DFR [18],

BSVQE [21], SINQ [20], and Lin's model [19]. Among them, BSVQE and SINQ are proposed not long ago. Note that BSQVE is a quality assessment algorithm for stereoscopic video, but we can treat the image as a video with only one frame and normally apply it to extract all the features. Fourthly, since the evaluated object is stereoscopic retargeted image, naturally, we consider to contrast with existing 2D retargeted image quality assessment metrics, due to image quality as the most noticed perceptual distortion dimension in 2D images. Here, we use HDPM [15], BNSSD [16], and MLF [17] put forward several months ago. What needs illustration is that MLF calls the Face++ API and do not provide more details about this feature, hence, other features except this feature are employed to compare. Finally, some widely used no-reference IQA metrics are utilized to test on SIRD database, which are fast EMD [12], SIFT flow [11], BRISQUE [13], and OG-IQA [14]. The reason why we do not consider 2D full-reference IQA metrics in the comparison is that some full-reference IQA

TABLE II
PERFORMANCE AND COMPARISON ON IEEE-SA DATABASE

| | Metric | PLCC | SRCC | KRCC | RMSE |
|---|---|---|---|---|---|
| 3D VCA | 3DAVM [3] | 0.8604 | 0.7831 | - | - |
| | Nojiri et al. [25] | 0.6854 | 0.6108 | - | 0.6088 |
| | Yano et al. [24] | 0.3988 | 0.3363 | - | 0.7602 |
| | Choi et al. [27] | 0.6509 | 0.5851 | - | 0.7096 |
| | Kim et al. [26] | 0.7018 | 0.6151 | - | 0.5838 |
| | Kim et al. [5] | **0.9042** | **0.8432** | - | **0.3491** |
| | Park et al. [4] | 0.8310 | 0.7534 | - | 0.4489 |
| Best SIRD | SRIQA [23] | * | * | * | * |
| | BSVQE [21] | 0.7052 | 0.6223 | 0.4564 | 0.5831 |
| | BNSSD [16] | * | * | * | * |
| Proposed | BRISQUE [13] | 0.7191 | 0.6508 | 0.4837 | 0.5765 |
| | DF+D-NSS | 0.8510 | 0.7667 | 0.5930 | 0.4243 |

VCA denotes visual comfort assessment. Best SIRD denotes the best metrics which perform best in their respective fields on SIRD database. DF denotes all the three disparity related features in the binocular incongruity measurement.

TABLE III
PERFORMANCE AND COMPARISON ON NBU 3D-VCA DATABASE

| | Metric | PLCC | SRCC | KRCC | RMSE |
|---|---|---|---|---|---|
| 3D VCA | Jung et al. [57] | 0.7784 | 0.7647 | - | 0.5035 |
| | Sohn et al. [6] | 0.7868 | 0.7608 | - | 0.4820 |
| | Jiang et al. [56] | 0.8078 | 0.7682 | - | 0.4623 |
| | Kim et al. [5] | **0.8128** | **0.7684** | - | **0.4015** |
| Proposed | DF+D-NSS | 0.8241 | 0.7870 | 0.6168 | 0.5577 |

VCA denotes visual comfort assessment. DF denotes all the three disparity related features in the binocular incongruity measurement.

TABLE IV
CONTRIBUTION OF EACH FEATURE COMPONENT ON SIRD DATABASE

| Metric | PLCC | SRCC | KRCC | RMSE |
|---|---|---|---|---|
| Local-SSIM+D-NSS+SD | 0.9101 | 0.8992 | 0.7277 | 0.4357 |
| DF+Local-SSIM+SD | 0.9296 | 0.9224 | 0.7632 | 0.3866 |
| DF+Local-SSIM+D-NSS | 0.9281 | 0.9213 | 0.7615 | 0.3893 |
| DF+D-NSS+SD | 0.9279 | 0.9207 | 0.7638 | 0.3902 |
| Proposed (All) | **0.9326** | **0.9260** | **0.7722** | **0.3772** |

DF denotes all the three disparity related features of the binocular incongruity measurement. And SD represents the semantic distortion feature.

metrics usually require the reference image and the distortion image are the same size, under the circumstances, the retargeted image and its original image do not satisfy the condition. In this way, we verify the performance of the proposed metric on SIRD database from all possible quality assessment angles. In each quality assessment area, the best performance is in bold for the sake of contrastive analysis.

From the results shown in Table I, we can observe that our proposed HDA-VCA metric obviously outperforms any other method, no matter in which quality assessment area. The PLCC and the SRCC of our metric can achieve 0.9326 and 0.9260, respectively. However, the second best performance is only 0.8789 in PLCC and 0.8573 in SRCC. The interval can reach 0.0537 at least.

*B. Performance and Comparison on IEEE-SA Database*

Except for stereoscopic retargeted images, we also try to make the proposed metric applicable to VCA for stereoscopic images in order to increase its universality in practical applications. Hence, the performance of our algorithm on IEEE-SA database is evaluated, which is mainly conducted from three perspectives.

Firstly, as the results presented in Table II, we compare with some typical VCA for stereoscopic image methods proposed on IEEE-SA database [3], [5], [24]-[27]. Among these methods, Kim's deep learning-based model [5] has reached the state-of-the-art performance. Secondly, the best metrics, which perform best in their respective fields on SIRD database as shown in Table I, are utilized to contrast with our metric. Park's model is the best 3D VCA algorithm on SIRD database. Note that because the IEEE-SA database is a purely stereoscopic image set rather than a SIR set, there are no reference images and distorted images, or source images and retargeted images. Thus the performance of SRIQA cannot be obtained which requires the retargeted image and its corresponding original image as inputs. For the same above-mentioned reason, BNSSD also demands the retargeted and original images as inputs and then cannot be verified on IEEE-SA database. Finally, in our metric, only the two kinds of no-reference measurements, i.e., the binocular incongruity measurement and the D-NSS measurement, are employed to test on IEEE-SA database.

It is not difficult to find from Table II that our proposed features can achieve the state-of-art performance for the VCA of general stereoscopic images, which is second only to 3DAVM model [3] and the deep learning-based model [5]. And the difference with 3DAVM is only 0.01 in PLCC, while the deep learning-based model relies on the training data, which may have been overfitting and not be suitable for VCA of stereoscopic retargeted images. By comparing the performance of our metric with that of all metric in Best SIRD section, although Park's model is proposed for VCA of general stereoscopic images, we can conclude that: the best metrics in their perspective field on SIRD database cannot effectively reflect the visual comfort of stereoscopic images simultaneously, while our metric can be applied in the VCA for both stereoscopic retargeted images and general stereo pairs with various disparity, and the latter can still achieve excellent performance.

*C. Performance and Comparison on NBU 3D-VCA Database*

To further verify the robustness of the propose HDA-VCA metric, expect for IEEE-SA database, we also conduct the experiment analysis on another public NBU 3D-VCA database. Meanwhile, as the experimental results presented in Table III, we compare our metric with four existing VCA algorithms [5], [6], [56], [57].

As can be seen from Table III, our proposed metric can reach the highest correlation on NBU 3D-VCA database. And one possible explanation for the higher RMSE value is that the absolute visual comfort evaluation scale of this database has some difference from SIRD and IEEE-SA databases, while the relative quality is more convincing in one database. Furthermore, comparing Table III with Table II, we can observe that Kim's model [5] has achieved the best performance on IEEE-SA database, while it performs worse than our metric on NBU 3D-VCA database. It further indicates that the deep learning-based VCA model indeed relies on the training data and it may have overfitting. While our metric analyzing the peculiarity and the binocular vision mechanism of HVS is more robust, whether it is used for VCA-SIR or VCA of generally stereoscopic image.

*D. Ablation Experiment*

To make the validity clearer, we further analyze the contribution of each measurement component. In detail, we drop out a kind of measurement each time to find out the improvement to the overall performance of our model. The results are given in Table IV.

From Table IV, it is easy to find that each proposed measurement component has the positive influence on the overall performance. Especially, the three disparity features in binocular incongruity measurement have the biggest contribution to the whole metric. And only the combination of four measurements can achieve the best performance.

V. CONCLUSION

In this paper, we have presented a hybrid distortion aggregated visual comfort assessment (HDA-VCA) metric for stereoscopic image retargeting (SIR), which is also the first VCA-SIR metric. Hybrid distortion types from structure distortion, information loss, binocular incongruity and semantic distortion are considered and evaluated to build final overall comfort assessment results. Comprehensive experiment results demonstrate that our proposed HDA-VCA metric has the remarkable ability to depict visual comfort of stereoscopic retargeted images compared with the state-of-the-art solutions.

In the future, we will expand our work to consider more impact factors into visual comfort assessment, including the 3D VR scenario which is also an important and challenging research topic.